\documentclass{cernrep}
\usepackage{graphicx,here} 
\usepackage{amsfonts,amsbsy}
\font\tenimbf=cmmib10 at 10pt
\font\sevenimbf=cmmib10 at 6pt
\font\fiveimbf=cmmib10 at 4pt
\newfam\imbf
\textfont\imbf=\tenimbf
\scriptfont\imbf=\sevenimbf
\scriptscriptfont\imbf=\fiveimbf

\begin{document}

\title{Hard scattering cross sections at LHC in the Glauber approach: from $pp$ to $pA$ and $AA$ collisions
\footnote{$^\star$ Contribution to the CERN Yellow Report on Hard Probes in Heavy Ion Collisions at the LHC}}
\author{David d'Enterria\footnote{$^\dagger$ Present address: Columbia University Nevis Laboratories, NY 10533, USA. 
[denterria$@$nevis.columbia.edu]}}
\institute{SUBATECH, BP 20722, 44307 Nantes Cedex 3, France.}

\maketitle

\begin{abstract}
The scaling rules of the invariant yields and cross sections for hard scattering
processes in proton-nucleus ($pA$) and nucleus-nucleus ($AB$) reactions at LHC energies relative to those of 
nucleon-nucleon $NN$ (isospin averaged $pp$) collisions are reviewed within the 
Glauber geometrical formalism. The number of binary $NN$ inelastic collisions for different 
centrality classes in p+Pb and Pb+Pb collisions at $\sqrt{s_{\mbox{\scriptsize{\it{NN}}}}}$ 
= 8.8 TeV and 5.5 TeV respectively, as obtained from a Glauber Monte Carlo, are also given.
\end{abstract}

\subsection{Proton-nucleus ($pA$) collisions}

\subsubsection{Glauber formalism}

The inelastic cross-section of a $p+A$ reaction, $\sigma_{pA}$, 
can be derived in the eikonal limit (straight line trajectories of colliding nucleons)
from the corresponding inelastic nucleon-nucleon $NN$ cross-section, 
$\sigma_{\mbox{\scriptsize{\it{NN}}}}(s)$ at the center-of-mass energy $\sqrt{s}$, and the geometry 
of the $pA$ collision simply determined by the impact parameter $b$ of the reaction. 
In the Glauber multiple collision model~\cite{enterria:glau}, such a cross-section reads 

\begin{equation}
\sigma_{pA}=\int d^2b \left[1-e^{-\sigma_{\mbox{\scriptsize{\it{NN}}}}(s)\,T_A(b)}\right]\; ,
\label{eq:glauber_pA}
\end{equation}

where $T_A(b)$ is the {\it nuclear thickness function} (or {\it nuclear profile function}) 
of the nucleus $A$ at impact parameter $b$:

\begin{equation}
T_A(b)=\int dz \;\rho_A(b,z).
\label{eq:nuc_profile}
\end{equation}

$T_A(b)$ gives the number of nucleons in the nucleus $A$ per unit area along a direction 
$z$ separated from the center of the nucleus by an impact parameter $b$. The nuclear
density, $\rho_A(b,z)$, is usually parametrized by a Woods-Saxon distribution with 
nuclear radius $R_A = 1.19\cdot A^{1/3} - 1.61\cdot A^{-1/3}$ fm and surface thickness $a=0.54$ fm 
as given by the experimental data \cite{enterria:deJae} and normalized so that

\begin{equation}
\int d^2b\; T_A(b) \;= \; A.
\label{eq:norm_T_A}
\end{equation}

\subsubsection{Hard scattering cross-sections}

Though Eq. (\ref{eq:glauber_pA}) is a general expression for the {\it total} inelastic cross-section, 
it can be applied to an inclusive $p+A\rightarrow h+X$ process of production of particle $h$. 
When one considers hard scattering processes, 
the corresponding cross-section $\sigma_{\mbox{\scriptsize{\it{NN}}}}^{hard}$ is small and one can 
expand Eq. (\ref{eq:glauber_pA}) in orders of $\sigma_{\mbox{\scriptsize{\it{NN}}}} T_A(b)$ and then, 
to first approximation 

\begin{equation}
\sigma_{pA}^{hard}\approx \int d^2b\; \sigma_{\mbox{\scriptsize{\it{NN}}}}^{hard}\;T_{A}(b)
\label{eq:glauber_pA_2}
\end{equation}

\subsubsection{``Minimum bias'' hard scattering cross-sections} 

Integrating Eq. (\ref{eq:glauber_pA_2}) over impact parameter, and using (\ref{eq:norm_T_A}),
one gets the {\it minimum bias} ($MB$) cross-section for a given hard process in $pA$ collisions
relative to the same cross-section in $pp$ (or $NN$) collisions:

\begin{equation}
(\sigma_{pA}^{hard})_{\mbox{\scriptsize{\it{MB}}}} = \;A\cdot\sigma_{\mbox{\scriptsize{\it{NN}}}}^{hard}
\label{eq:glauber_pA_minbias}
\end{equation}


From this expression it is easy to see that the corresponding 
minimum bias multiplicity (invariant yield per nuclear reaction: 
$N^{hard}_{\mbox{\scriptsize{\it{NN,pA}}}}=\sigma^{hard}_{\mbox{\scriptsize{\it{NN,pA}}}}/\sigma_{\mbox{\scriptsize{\it{NN,pA}}}}^{geo}$) for a given hard-process in a $pA$ collision compared to that of a $pp$ collision is

\begin{equation}
\langle N_{\mbox{\scriptsize{\it{pA}}}}^{hard}\rangle_{\mbox{\scriptsize{\it{MB}}}} = \;A\cdot\frac{\sigma_{\mbox{\scriptsize{\it{NN}}}}}{\sigma_{\mbox{\scriptsize{\it{pA}}}}^{geo}}\cdot N_{\mbox{\scriptsize{\it{NN}}}}^{hard} = \frac{A}{\sigma_{\mbox{\scriptsize{\it{pA}}}}^{geo}}\cdot\sigma_{\mbox{\scriptsize{\it{NN}}}}^{hard}\; ,
\label{eq:Nhard_pA_minbias}
\end{equation}

where $\sigma_{\mbox{\scriptsize{\it{pA}}}}^{geo}$ is the geometrical 
$pA$ cross-section given, in its most general form, by Eq. (\ref{eq:glauber_pA}). 
The average nuclear thickness function for {\it minimum bias} reactions 
[making use of Eq. (\ref{eq:norm_T_A})] reads:

\begin{equation}
\langle T_{\mbox{\scriptsize{\it{A}}}}\rangle_{\mbox{\scriptsize{\it{MB}}}} \equiv \frac{\int d^2b \; T_{\mbox{\scriptsize{\it{A}}}}}{\int d^2b} \;= \frac{A}{\pi \; R_A^2}=\frac{A}{\sigma_{\mbox{\scriptsize{\it{pA}}}}^{geo}}.
\label{eq:T_pA_minbias}
\end{equation}

Thus, for a p+Pb ($A(Pb)$ = 208) collision at LHC energies $\sqrt{s_{\mbox{\scriptsize{\it{NN}}}}}$ = 8.8 TeV with

\begin{eqnarray}
\sigma_{\mbox{\scriptsize{\it{NN}}}} & \approx & 77 \; \mbox{ mb \cite{NNcross-section}, and } \\
\sigma_{\mbox{\scriptsize{\it{pPb}}}}^{geo} & \approx & 2162 \; \mbox{ mb \cite{pPbcross-section},} 
\end{eqnarray}

one obtains: 
$\langle N_{\mbox{\scriptsize{\it{pPb}}}}^{hard}\rangle_{\mbox{\scriptsize{\it{MB}}}} \approx 7.4
\cdot N_{\mbox{\scriptsize{\it{NN}}}}^{hard}$, and the average nuclear thickness
function amounts to $\langle T_{\mbox{\scriptsize{\it{Pb}}}}\rangle_{\mbox{\scriptsize{\it{MB}}}}\; 
=$ 0.096 mb$^{-1}$ = 0.96 fm$^{-2}$.\\

\subsection{Nucleus-nucleus ($AB$) collisions}

\subsubsection{Glauber formalism}

As in the proton-nucleus case, the inclusive inelastic cross-section $\sigma_{\mbox{\scriptsize{\it{AB}}}}$ 
for a collision of nuclei $A$ and $B$ is given in the multiple-scattering Glauber approximation by:

\begin{equation}
\sigma_{\mbox{\scriptsize{\it{AB}}}}=\int d^2b \left[1-e^{-\sigma_{\mbox{\scriptsize{\it{NN}}}}(s)\,T_{\mbox{\scriptsize{\it{AB}}}}(b)}\right]\; ,
\label{eq:glauber_AB}
\end{equation}

where now $T_{\mbox{\scriptsize{\it{AB}}}}(b)$ is the {\it nuclear overlap function} of the nuclei $A$ and $B$ separated
by impact parameter $b$. $T_{\mbox{\scriptsize{\it{AB}}}}(b)$ can be written as a convolution of the corresponding
thickness functions of $A$ and $B$ over the element of overlapping area $d^2{\vec s}$
 ($\vec{s}=(x,y)$ is a 2-D vector in the transverse plane, and $\vec{b}$ is the impact parameter 
between the centers of the nuclei):

\begin{equation}
T_{\mbox{\scriptsize{\it{AB}}}}(b)\; = \;\int d^2{\vec s} \;T_A({\vec s})\;T_B(|{\vec b}-{\vec s}|).
\label{eq:nuc_overlap}
\end{equation}

$T_{\mbox{\scriptsize{\it{AB}}}}(b)$ is normalized so that integrating over all impact parameters one gets:

\begin{equation}
\int d^2b\; T_{\mbox{\scriptsize{\it{AB}}}}(b)= A\,B.
\label{eq:norm_T_AB}
\end{equation}

\subsubsection{Hard scattering cross-sections}

As in the $pA$ case, for hard processes of the type $A+B\rightarrow h+X$, Eq. (\ref{eq:glauber_AB}), 
can be approximated by:

\begin{equation}
\sigma_{\mbox{\scriptsize{\it{AB}}}}^{hard}\approx \int d^2b\; \sigma_{\mbox{\scriptsize{\it{NN}}}}^{hard}\;T_{\mbox{\scriptsize{\it{AB}}}}(b).
\label{eq:glauber_AB_2}
\end{equation}

\subsubsection{``Minimum bias'' hard scattering cross-sections and yields}

Integrating Eq. (\ref{eq:glauber_AB_2}) over impact parameter and using (\ref{eq:norm_T_AB}),
one gets the {\it minimum bias} ($MB$) cross-section for a given hard process in $AB$ collisions
relative to the corresponding $pp$ cross-section:

\begin{equation}
(\sigma_{\mbox{\scriptsize{\it{AB}}}}^{hard})_{\mbox{\scriptsize{\it{MB}}}} = \;A\cdot B\cdot\sigma_{\mbox{\scriptsize{\it{NN}}}}^{hard}
\label{eq:sigma_glauber_AB_minbias}
\end{equation}

Again the corresponding {\it minimum bias} multiplicity (invariant yield per nuclear reaction: 
$N^{hard}_{\mbox{\scriptsize{\it{NN,AB}}}}=\sigma^{hard}_{\mbox{\scriptsize{\it{NN,AB}}}}/\sigma_{\mbox{\scriptsize{\it{NN,AB}}}}^{geo}$) for a given hard-process in a $AB$ collision compared to that of a $pp$ collision is

\begin{equation}
\langle N_{\mbox{\scriptsize{\it{AB}}}}^{hard}\rangle_{\mbox{\scriptsize{\it{MB}}}} = \;A\cdot B\cdot\frac{\sigma_{\mbox{\scriptsize{\it{NN}}}}}{\sigma_{\mbox{\scriptsize{\it{AB}}}}^{geo}}\cdot N_{\mbox{\scriptsize{\it{NN}}}}^{hard} = \frac{A\cdot B}{\sigma_{\mbox{\scriptsize{\it{AB}}}}^{geo}}\cdot\sigma_{\mbox{\scriptsize{\it{NN}}}}^{hard}\; ,
\label{eq:Nhard_AB_minbias}
\end{equation}

where $\sigma_{\mbox{\scriptsize{\it{AB}}}}^{geo}$ is the geometrical 
$AB$ cross-section given, in its most general form, by Eq. (\ref{eq:glauber_AB}). 
The average nuclear overlap function for {\it minimum bias} reactions 
[making use of Eq. (\ref{eq:norm_T_AB})] reads now:

\begin{equation}
\langle T_{\mbox{\scriptsize{\it{AB}}}}\rangle_{\mbox{\scriptsize{\it{MB}}}} \equiv \frac{\int d^2b \; T_{\mbox{\scriptsize{\it{AB}}}}}{\int d^2b} \;= \frac{A\cdot B}{\pi (R_A+R_B)^2}=\frac{AB}{\sigma_{\mbox{\scriptsize{\it{AB}}}}^{geo}},
\label{eq:T_AB_minbias}
\end{equation}

Thus, for a Pb+Pb ($A^2(Pb)$ = 43264) collision at LHC energies $\sqrt{s_{\mbox{\scriptsize{\it{NN}}}}}$ = 5.5 TeV with

\begin{eqnarray}
\sigma_{\mbox{\scriptsize{\it{NN}}}} & \approx & 72 \; \mbox{ mb \cite{NNcross-section}, and } \\
\sigma_{\mbox{\scriptsize{\it{PbPb}}}}^{geo} & \approx & 7745 \; \mbox{ mb \cite{PbPbcross-section},} 
\end{eqnarray}

one gets: 
$\langle N_{\mbox{\scriptsize{\it{PbPb}}}}^{hard}\rangle_{\mbox{\scriptsize{\it{MB}}}} \approx 400
\cdot N_{\mbox{\scriptsize{\it{NN}}}}^{hard}$, and the average nuclear overlap
function amounts to $\langle T_{\mbox{\scriptsize{\it{PbPb}}}}\rangle_{\mbox{\scriptsize{\it{MB}}}}\; 
=$ 5.58 mb$^{-1}$ = 55.8 fm$^{-2}$.\\

\subsubsection{Binary collision scaling}

For a given impact parameter $b$ the {\it average} hard scattering yield can be obtained by 
multiplying each nucleon in nucleus $A$ against the density it sees along the 
$z$ direction in nucleus $B$, then integrated over all of nucleus $A$, i.e.

\begin{equation}
\langle N_{\mbox{\scriptsize{\it{AB}}}}^{hard}\rangle (b) \;=\; 
\sigma_{\mbox{\scriptsize{\it{NN}}}}^{hard}\;\int d^2{\vec s} \;\int \rho_A({\vec s,z'})\;\int \rho_B(|{\vec b}-{\vec s}|,z'')\;dz''dz'\;
\equiv \; \sigma_{\mbox{\scriptsize{\it{NN}}}}^{hard}\cdot T_{\mbox{\scriptsize{\it{AB}}}}(b)\;,
\label{eq:N_AB}
\end{equation}

where we have made use of expressions (\ref{eq:nuc_profile}) and (\ref{eq:nuc_overlap}).
In the same way, one can obtain a useful expression for the probability
of an inelastic $NN$ collision or, equivalently, for the {\it average} number of 
binary inelastic collisions, $\langle N_{coll} \rangle$, in a nucleus-nucleus reaction
with impact parameter $b$:

\begin{equation}
\langle N_{coll}\rangle (b) \; = \; \sigma_{\mbox{\scriptsize{\it{NN}}}}\cdot T_{\mbox{\scriptsize{\it{AB}}}}(b)
\label{eq:N_coll}
\end{equation}

From this last expression one can see that the nuclear overlap function, 
$T_{\mbox{\scriptsize{\it{AB}}}}(b)=N_{coll}(b)/\sigma_{\mbox{\scriptsize{\it{NN}}}}$ 
[mb$^{-1}$], can be thought as the luminosity (reaction rate per unit of cross-section) per $AB$ collision 
at a given impact parameter. As an example, the average number of binary collisions in minimum bias 
Pb+Pb reactions at LHC ($\sigma_{\mbox{\scriptsize{\it{NN}}}}$ = 72 mb = 7.2 fm$^{2}$) is:

\begin{equation}
\langle N_{coll}\rangle_{\mbox{\scriptsize{\it{MB}}}} = 7.2\;\mbox{fm}^{2}\;\cdot\;55.9\;\mbox{fm}^{-2}\;=\;400.
\end{equation}

From (\ref{eq:N_AB}) and (\ref{eq:N_coll}), we get so-called ``binary (or point-like) scaling'' 
formula for the hard scattering yields in heavy-ion reactions:

\begin{equation}
\langle N_{\mbox{\scriptsize{\it{AB}}}}^{hard}\rangle (b) \approx 
\langle N_{coll}\rangle (b) \cdot N_{\mbox{\scriptsize{\it{NN}}}}^{hard}
\label{eq:binary_scaling}
\end{equation}

\subsubsection{Hard scattering yields and cross-sections in a given centrality class}

Equation (\ref{eq:glauber_AB_2}) gives the reaction cross-section for a given hard process 
in $AB$ collisions {\it at a given impact parameter} $b$ as a function of the corresponding
reaction cross-section in $pp$ collisions. Very usually, however, in nucleus-nucleus collisions we 
are interested in calculating such a reaction cross-section for a given {\it centrality class}, 
$(\sigma_{AB}^{hard})_{C_1-C_2}$, where the centrality selection 
$C_1-C_2$ corresponds to integrating Eq. (\ref{eq:glauber_AB_2}) between impact parameters $b_1$ and $b_2$.
It is useful, in this case, to define two parameters \cite{enterria:vogt,enterria:star}:

\begin{itemize}
\item The fraction of the total cross-section for hard processes occurring at impact parameters
$b_{1}<b<b_{2}$  ($d^2b = 2\pi bdb$):
\begin{equation}
f_{hard}(b_1<b<b_2)\;=\;\frac{2\pi}{AB}\int_{b_1}^{b_2} bdb\; T_{\mbox{\scriptsize{\it{AB}}}}(b).
\label{eq:f_AB}
\end{equation}
\item The fraction of the geometric cross-section with impact parameter $b_{1}<b<b_{2}$:
\begin{equation}
f_{geo}(b_1<b<b_2)\;=\;\left [2\pi\; \int_{b_1}^{b_2} bdb \left(1-e^{-\sigma_{\mbox{\scriptsize{\it{NN}}}}
T_{\mbox{\scriptsize{\it{AB}}}}(b)}\right)\right]/\sigma_{\mbox{\scriptsize{\it{AB}}}}^{geo}\; ,
\label{eq:f_geo}
\end{equation}
[$f_{geo}$ simply corresponds to a 0.$X$ (e.g. 0.1) factor for the $X$\%(10\%) centrality.]
\end{itemize}

Hard scattering production is more enhanced for increasingly central reactions 
(with larger number on $N_{coll}$) as compared to the total reaction cross-section 
(which includes ``soft'', - scaling with the number of participant nucleons $N_{part}$ -, 
as well as ``hard'' contributions). The growth with $b$ of the geometric cross-section is 
slower than that of the hard component. For this reason, the behaviour 
of $f_{\mbox{\scriptsize{\it{hard}}}}$ and $f_{geo}$ as a function of $b$, 
although similar in shape is not the same (see \cite{enterria:vogt}): 
$f_{hard}\approx$ 1 for $b$ = 2$R_A$, but $f_{geo}\approx$ 0.75 for $b$ = 2$R_A$.\\

Similarly to (\ref{eq:T_AB_minbias}), we can obtain now the nuclear overlap function for 
any given centrality class $C_1-C_2$:

\begin{equation}
\langle T_{\mbox{\scriptsize{\it{AB}}}}\rangle_{C_1-C_2} \;\equiv\; \frac{\int^{b_2}_{b_1} d^2b \; T_{\mbox{\scriptsize{\it{AB}}}}}
{\int^{b_2}_{b_1} d^2b} = \frac{A\cdot B}{\sigma_{\mbox{\scriptsize{\it{AB}}}}^{geo}} \cdot \frac{f_{hard}}{f_{geo}}\;
\label{eq:T_AB_centrality}
\end{equation}

The number of hard processes per nuclear collision for reactions with
impact parameter $b_1<b<b_2$ is given by

\begin{equation}
\langle N_{\mbox{\scriptsize{\it{AB}}}}^{\mbox{\scriptsize{\it{hard}}}}\rangle_{C_1-C_2} = 
\frac{\sigma_{\mbox{\scriptsize{\it{AB}}}}^{hard}\mbox{\small{$(b_1<b<b_2$)}}}
{\sigma_{\mbox{\scriptsize{\it{AB}}}}^{geo}\mbox{\small{$(b_1<b<b_2)$}}} = A\cdot B\cdot \frac{\sigma_{\mbox{\scriptsize{\it{NN}}}}
^{\mbox{\scriptsize{\it{hard}}}}}{\sigma_{\mbox{\scriptsize{\it{AB}}}}^{geo}} \cdot \frac{f_{hard}}{f_{geo}} \; ,
\label{eq:N_hard_centrality}
\end{equation}

which we could have just obtained directly from (\ref{eq:N_AB}) and (\ref{eq:T_AB_centrality}). From 
(\ref{eq:Nhard_AB_minbias}) and (\ref{eq:N_hard_centrality}) it is also easy to see that:

\begin{equation}
\langle N_{\mbox{\scriptsize{\it{AB}}}}^{\mbox{\scriptsize{\it{hard}}}}\rangle_{C_1-C_2} = 
\langle N_{\mbox{\scriptsize{\it{AB}}}}^{\mbox{\scriptsize{\it{hard}}}}\rangle_{\mbox{\scriptsize{\it{MB}}}} \cdot \frac{f_{hard}}{f_{geo}}
\label{eq:N_hard_centrality2}
\end{equation}

Finally, the cross-section for hard processes produced in the centrality class $C_1-C_2$ (corresponding
to a fraction $f_{geo}$ of the reaction cross-section, $(\sigma_{\mbox{\scriptsize{\it{AB}}}}^{geo})_{C_1-C_2}$) 
is:

\begin{equation}
(\sigma_{\mbox{\scriptsize{\it{AB}}}}^{\mbox{\scriptsize{\it{hard}}}})_{C_1-C_2} = 
A\cdot B \cdot f_{hard} \cdot \sigma_{\mbox{\scriptsize{\it{NN}}}}^{\mbox{\scriptsize{\it{hard}}}}
\label{eq:sigma_hard_centrality}
\end{equation}

\begin{figure}[htbp]
\begin{center}
\includegraphics[height=7.0cm]{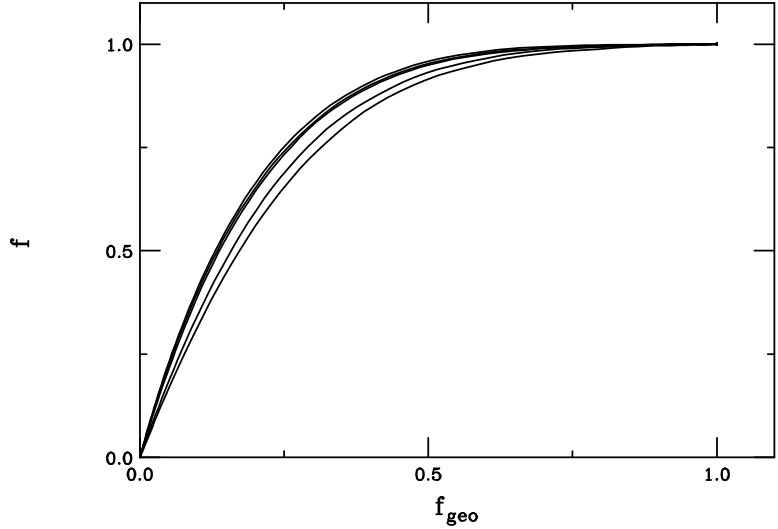}
\end{center}
\caption{\it Figure 4 of ref. \protect\cite{enterria:vogt}. Fraction of the hard cross-section, 
$f\equiv f_{hard}(0<b<b_2)$, vs. the fraction of the total geometrical cross-section, 
$f_{geo}(0<b<b_1)$, for several heavy-ion collisions (from left to right): 
197+197, 110+197, 63+193, 27+197, and 16+197.}
\label{fig:fgeo_vs_fhard}
\end{figure}

Figure \ref{fig:fgeo_vs_fhard}, extracted from \cite{enterria:vogt}, plots the (top) fraction of the hard 
cross-section, $f_{hard}(0<b<b_2)$ (labeled in the plot as $f$), as a function of 
the top fraction of the total geometrical cross-section, $f_{geo}(0<b<b_2)$,
for several nucleus-nucleus reactions.

As a practical application of Eq. (\ref{eq:sigma_hard_centrality}) and the results of Fig. \ref{fig:fgeo_vs_fhard}, 
the hard-scattering cross-sections in Pb+Pb for the top 0-10\% ($f_{hard}$ = 0.41 for $f_{geo}$ = 0.1
from the practically equivalent Au+Au system of figure \ref{fig:fgeo_vs_fhard}) 
and 0-20\% ($f_{hard}$ = 0.664 for $f_{geo}$ = 0.2) central collisions relate to the $pp$ cross-section, 
in the absence of nuclear effects, respectively as:

\begin{eqnarray}
(\sigma_{\mbox{\scriptsize{\it{AB}}}}^{\mbox{\scriptsize{\it{hard}}}})_{0-10\%} & = & (208)^2\cdot 0.41\cdot \sigma_{\mbox{\scriptsize{\it{NN}}}}^{\mbox{\scriptsize{\it{hard}}}}\;\approx\;1.7\cdot 10^{4}\cdot\sigma_{\mbox{\scriptsize{\it{NN}}}}^{\mbox{\scriptsize{\it{hard}}}}\\
(\sigma_{\mbox{\scriptsize{\it{AB}}}}^{\mbox{\scriptsize{\it{hard}}}})_{0-20\%} & = & (208)^2\cdot 0.664 \cdot \sigma_{\mbox{\scriptsize{\it{NN}}}}^{\mbox{\scriptsize{\it{hard}}}}\;\approx\;2.9\cdot 10^{4}\cdot\sigma_{\mbox{\scriptsize{\it{NN}}}}^{\mbox{\scriptsize{\it{hard}}}}
\label{eq:N_hard_values}
\end{eqnarray}

\begin{center}
\line(1,0){100}
\end{center}

A straightforward way to compute the invariant yield for a given hard 
process in a given centrality class of a nucleus-nucleus collision from the corresponding yield 
in $pp$ collisions consists in determining, via a Glauber MC calculation, the average number of inelastic 
$NN$ collisions corresponding to that centrality class via 

\begin{equation}
\langle N_{coll}\rangle_{C_1-C_2} \; =\; \langle T_{\mbox{\scriptsize{\it{AB}}}}\rangle_{C_1-C_2} \cdot 
\sigma_{\mbox{\scriptsize{\it{NN}}}}\; ,
\end{equation}

and then use this value in the ``binary-scaling'' formula 

\begin{eqnarray}
\langle N_{\mbox{\scriptsize{\it{AB}}}}^{hard}\rangle_{C_1-C_2} & = &
\langle N_{coll}\rangle_{C_1-C_2} \cdot N_{\mbox{\scriptsize{\it{NN}}}}^{hard}\;,\;\;\;\;\;\mbox{ or}\\
\label{eq:binary_scaling_yields1}
\nonumber\\
(\sigma_{\mbox{\scriptsize{\it{AB}}}}^{\mbox{\scriptsize{\it{hard}}}})_{C_1-C_2}  & = & \langle T_{AB}\rangle_{C_1-C_2} \cdot 
\,(\sigma_{\mbox{\scriptsize{\it{AB}}}}^{geo})_{C_1-C_2}\,\cdot\sigma_{\mbox{\scriptsize{\it{NN}}}}^{\mbox{\scriptsize{\it{hard}}}}
\label{eq:binary_scaling_yields2}
\end{eqnarray}

The same two formulae above apply to $pA$ collisions (of course substituting $AB$ by $pA$ and computing
$N_{coll}$ from $T_{A}$ instead of from $T_{AB}$).\\

Finally, to obtain the {\it experimental rates}, 
$({\cal N}_{\mbox{\scriptsize{\it{AB}}}}^{\mbox{\scriptsize{\it{hard}}}})_{C_1-C_2}$, 
actually measured in a given centrality bin one needs to take into account the expected 
{\it integrated luminosity} ${\cal L}_{int}$ [mb$^{-1}$] as follows:

\begin{eqnarray}
({\cal N}_{\mbox{\scriptsize{\it{AB}}}}^{\mbox{\scriptsize{\it{hard}}}})_{C_1-C_2} & = & {\cal L}_{int} \cdot 
(\sigma_{\mbox{\scriptsize{\it{AB}}}}^{\mbox{\scriptsize{\it{hard}}}})_{C_1-C_2} 
\label{eq:experimental_rates}
\end{eqnarray}

\subsection{Hard scattering yields and cross-sections for $pPb$ and $PbPb$ collisions at LHC}

As a practical application of the Glauber approach described here, in Table
\ref{tab:glauber_Ncoll}, the values of $\langle N_{coll}\rangle$ and $\langle T_{pPb} \rangle$ 
are quoted for different centrality classes obtained from a Monte Carlo 
calculation \cite{enterria:klaus} for p+Pb ($\sqrt{s_{\mbox{\scriptsize{\it{NN}}}}}$ = 8.8 TeV 
and $\sigma_{\mbox{\scriptsize{\it{NN}}}}$ = 77 mb) and Pb+Pb 
($\sqrt{s_{\mbox{\scriptsize{\it{NN}}}}}$ = 5.5 TeV for an inelastic $pp$ cross-section 
of $\sigma_{\mbox{\scriptsize{\it{NN}}}}$ = 72 mb) collisions 
(Woods-Saxon Pb density parametrization with $R_A$ = 6.78 fm 
and $a$ = 0.54 fm).

\begin{table}[htb]
\begin{center}
\caption{Number of inelastic $NN$ collisions, $\langle N_{coll}\rangle$, 
and nuclear thickness $\langle T_{pPb} \rangle$ or overlap $\langle T_{PbPb} \rangle$ function 
per centrality class, in p+Pb ($\sqrt{s_{\mbox{\scriptsize{\it{NN}}}}}$ = 8.8 TeV, 
$\sigma_{\mbox{\scriptsize{\it{NN}}}}$ = 77 mb) and Pb+Pb collisions at LHC 
($\sqrt{s_{\mbox{\scriptsize{\it{NN}}}}}$ = 5.5 TeV, $\sigma_{\mbox{\scriptsize{\it{NN}}}}$ = 72 mb) 
obtained with the Glauber Monte Carlo code of ref. \protect\cite{enterria:klaus}. The errors
in $\langle N_{coll}\rangle$, not shown, are of the same order as the current uncertainty in the 
value of the nucleon-nucleon inelastic cross section, $\sigma_{\mbox{\scriptsize{\it{NN}}}}$, at
LHC energies ($\sim$ 10\%).}
\vskip0.4cm
\begin{tabular}{c|c|c|c|c}
\hline\hline
\hspace{1mm} 
Centrality ($C_1-C_2$)\hspace{1mm} & \multicolumn{2}{c|}{p+Pb}  & \multicolumn{2}{c}{Pb+Pb} \\
 & \hspace{6mm} $\langle N_{coll} \rangle$ \hspace{6mm} & \hspace{1mm} $\langle T_{pPb} \rangle$ (mb$^{-1}$) \hspace{1mm} & 
   \hspace{6mm} $\langle N_{coll} \rangle$ \hspace{7mm} & \hspace{1mm} $\langle T_{PbPb} \rangle$ (mb$^{-1}$) \hspace{1mm} \\\hline
\hline
  0- 5\% & 15.7 & 0.203                & 1876.0  & 26.0\\
  0-10\% & 15.3 & 0.198                & 1670.2  & 23.2 \\
 10-20\% & 13.8 & 0.179                & 1019.5  & 14.2 \\
 20-30\% & 12.0 & 0.155                &  612.4  &  8.50 \\
 30-40\% &  9.9 & 0.128                &  351.8  &  4.89 \\
 40-50\% &  7.8 & 0.101                &  188.0  &  2.61 \\
 50-60\% &  5.6 & 7.27$\cdot$10$^{-2}$ &   92.9  &  1.29 \\
 60-70\% &  3.8 & 4.93$\cdot$10$^{-2}$ &   41.4  &  5.75$\cdot$10$^{-1}$ \\
 70-80\% &  2.6 & 3.37$\cdot$10$^{-2}$ &   16.8  &  2.33$\cdot$10$^{-1}$ \\
 80-90\% &  1.7 & 2.20$\cdot$10$^{-2}$ &    6.7  &  9.31$\cdot$10$^{-2}$ \\
90-100\% &  1.2 & 1.55$\cdot$10$^{-2}$ &    2.7  &  3.75$\cdot$10$^{-2}$ \\
min. bias&  7.4 & 9.61$\cdot$10$^{-2}$ &  400.0  &  5.58\\\hline\hline
\end{tabular}
\label{tab:glauber_Ncoll}
\end{center}
\end{table}

Using (\ref{eq:binary_scaling_yields1}), (\ref{eq:binary_scaling_yields2}) and Table I, 
we can now easily get the scaling factors of the 
cross-sections and yields from $pp$ to, e.g., central (0-10\%), minimum bias, 
and semi-peripheral (60-80\%, from the combined average 60-70\% and 70-80\%) p+Pb (8.8 TeV) 
and Pb+Pb (5.5 TeV) collisions :\\

\newpage
For p+Pb collisions ($\sigma_{\mbox{\scriptsize{\it{pPb}}}}^{geo}$ = 2162 mb):

\begin{eqnarray}
\langle N_{\mbox{\scriptsize{\it{pPb}}}}^{\mbox{\scriptsize{\it{hard}}}}\rangle_{0-10\%}  = 15.3 \cdot N_{\mbox{\scriptsize{\it{NN}}}}^{hard} 
& \, , \,  & 
(\sigma_{\mbox{\scriptsize{\it{pPb}}}}^{\mbox{\scriptsize{\it{hard}}}})_{0-10\%} = 0.198 \cdot 0.1 \cdot 2162\cdot\sigma_{\mbox{\scriptsize{\it{NN}}}}^{\mbox{\scriptsize{\it{hard}}}} \;\approx\; 450 \cdot\sigma_{\mbox{\scriptsize{\it{NN}}}}^{\mbox{\scriptsize{\it{hard}}}}\\
\langle N_{\mbox{\scriptsize{\it{pPb}}}}^{\mbox{\scriptsize{\it{hard}}}}\rangle_{60-80\%}  = 3.2 \cdot N_{\mbox{\scriptsize{\it{NN}}}}^{hard} 
& \, , \,  & 
(\sigma_{\mbox{\scriptsize{\it{pPb}}}}^{\mbox{\scriptsize{\it{hard}}}})_{60-80\%} = 0.042 \cdot 0.2 \cdot 2162 \cdot\sigma_{\mbox{\scriptsize{\it{NN}}}}^{\mbox{\scriptsize{\it{hard}}}} \;\approx\; 18 \cdot\sigma_{\mbox{\scriptsize{\it{NN}}}}^{\mbox{\scriptsize{\it{hard}}}}\\
\langle N_{\mbox{\scriptsize{\it{pPb}}}}^{\mbox{\scriptsize{\it{hard}}}}\rangle_{MB} = 7.4 \cdot N_{\mbox{\scriptsize{\it{NN}}}}^{hard} 
& \, , \,  & 
(\sigma_{\mbox{\scriptsize{\it{pPb}}}}^{\mbox{\scriptsize{\it{hard}}}})_{MB} = 0.096 \cdot 2162 \cdot\sigma_{\mbox{\scriptsize{\it{NN}}}}^{\mbox{\scriptsize{\it{hard}}}} \;\approx\; 2\cdot 10^{2}\cdot\sigma_{\mbox{\scriptsize{\it{NN}}}}^{\mbox{\scriptsize{\it{hard}}}}
\label{eq:N_hard_pPb}
\end{eqnarray}

For Pb+Pb collisions ($\sigma_{\mbox{\scriptsize{\it{PbPb}}}}^{geo}$ = 7745 mb):

\begin{eqnarray}
\langle N_{\mbox{\scriptsize{\it{PbPb}}}}^{\mbox{\scriptsize{\it{hard}}}}\rangle_{0-10\%}  = 1670 \cdot N_{\mbox{\scriptsize{\it{NN}}}}^{hard} & \, , \,  & 
(\sigma_{\mbox{\scriptsize{\it{PbPb}}}}^{\mbox{\scriptsize{\it{hard}}}})_{0-10\%} = 23.2 \cdot 0.1 \cdot 7745 \cdot\sigma_{\mbox{\scriptsize{\it{NN}}}}^{\mbox{\scriptsize{\it{hard}}}} \approx 1.6\cdot 10^{4}\cdot\sigma_{\mbox{\scriptsize{\it{NN}}}}^{\mbox{\scriptsize{\it{hard}}}}\\
\langle N_{\mbox{\scriptsize{\it{PbPb}}}}^{\mbox{\scriptsize{\it{hard}}}}\rangle_{60-80\%} = 29.1 \cdot N_{\mbox{\scriptsize{\it{NN}}}}^{hard} & \, , \, & 
(\sigma_{\mbox{\scriptsize{\it{PbPb}}}}^{\mbox{\scriptsize{\it{hard}}}})_{60-80\%} = 0.4 \cdot 0.2 \cdot 7745\cdot\sigma_{\mbox{\scriptsize{\it{NN}}}}^{\mbox{\scriptsize{\it{hard}}}} \approx 6.2\cdot 10^{2}\cdot\sigma_{\mbox{\scriptsize{\it{NN}}}}^{\mbox{\scriptsize{\it{hard}}}}\\
\langle N_{\mbox{\scriptsize{\it{PbPb}}}}^{\mbox{\scriptsize{\it{hard}}}}\rangle_{MB} = 400 \cdot N_{\mbox{\scriptsize{\it{NN}}}}^{hard} & \, , \, & 
(\sigma_{\mbox{\scriptsize{\it{PbPb}}}}^{\mbox{\scriptsize{\it{hard}}}})_{MB} = 5.58 \cdot 7745\cdot\sigma_{\mbox{\scriptsize{\it{NN}}}}^{\mbox{\scriptsize{\it{hard}}}} \approx 4.3\cdot 10^{4}\cdot\sigma_{\mbox{\scriptsize{\it{NN}}}}^{\mbox{\scriptsize{\it{hard}}}}
\label{eq:N_hard_valuesPbPb}
\end{eqnarray}

\subsection{Nuclear effects in $p A$ and $AB$ collisions}

Eqs. (\ref{eq:glauber_pA_2}) and (\ref{eq:glauber_AB_2}) for the hard scattering cross-sections in $pA$ and $AB$ 
collisions have been derived within an eikonal framework which only takes into account the geometric aspects 
of the reactions. Any differences of the experimentally measured $\sigma_{pA,AB}^{hard}$ with respect to these 
expressions 
indicate ``de facto'' the existence of ``nuclear effects'' (such as e.g. ``shadowing'', ``Cronin enhancement'',
or ``parton energy loss'') not accounted for by the Glauber formalism. Indeed, in the multiple-scattering Glauber 
model each nucleon-nucleon collision is treated incoherently and thus, unaffected by any other scattering 
taking place before (initial-state) or after (final-state effects) it.\\

If the Glauber approximation holds, from (\ref{eq:glauber_pA_2}) and (\ref{eq:glauber_AB_2}) one 
would expect a $\propto A^1$, and $\propto A^2$ growth of the hard processes cross-section with system size 
respectively. Equivalently, since 
$N^{hard}_{\mbox{\scriptsize{\it{NN,AB}}}}=\sigma^{hard}_{\mbox{\scriptsize{\it{NN,AB}}}}/\sigma_{\mbox{\scriptsize{\it{NN,AA}}}}^{geo}$ and $\sigma_{\mbox{\scriptsize{\it{NN,AB}}}}^{geo}\sim R_A^2$ with $R_A\sim A^{1/3}$, one would expect a
growth of the {\it number} of hard process as $\propto A^{1/3}, \propto A^{4/3}$ for $pA, AA$ collisions respectively.
Experimentally, in minimum bias $pA$ and $AB$ collisions, it has been found that the 
production cross-sections for hard processes actually grow as:

\begin{equation}
(\sigma_{pA}^{hard})_{\mbox{\scriptsize{\it{MB}}}} = \;A^\alpha\cdot\sigma_{\mbox{\scriptsize{\it{NN}}}}^{hard} \;,\;\;\;\;
\mbox{ and }\;\;\; 
(\sigma_{\mbox{\scriptsize{\it{AB}}}}^{hard})_{\mbox{\scriptsize{\it{MB}}}} = \;(AB)^\alpha\cdot\sigma_{\mbox{\scriptsize{\it{NN}}}}^{hard} \;,\;\;\;\; \mbox{ with } \alpha\neq 1
\label{eq:minbias_exp}
\end{equation}

More precisely, in high-$p_T$ processes in $pA$ and heavy-ion collisions at SPS energies one founds $\alpha>$ 1 
(due to initial-state p$_T$ broadening or ``Cronin enhancement''); whereas $\alpha<$ 1 at 
RHIC energies (``high-$p_T$ suppression''). Theoretically,
one can still make predictions on the hard probe yields in $pA,AB$ collisions using the pQCD 
factorization machinery for the $pp$ cross-section complemented with the Glauber formalism while
modifying effectively the nuclear PDFs and parton fragmentation functions 
to take into account any initial- and/or final- state nuclear medium effect.\\

\subsection{Summary of useful formulae}

Finally, let us summarize a few useful formulae derived here to determine the hard-scattering 
invariant yields, cross-sections, or experimental rates, from $pp$ to $pA$ and $AB$ 
collisions for centrality bin $C_1-C_2$ (corresponding to a nuclear thickness $T_{A}$ or nuclear 
overlap function $T_{AB}$ and to an average number of $NN$ inelastic collisions $\langle N_{coll}\rangle$):

\begin{eqnarray}
\frac{(d^2 N_{\mbox{\tiny{\it{pA,AB}}}}^{hard})_{C_1-C_2}}{dp_Tdy} & = &
\langle T_{A,AB}\rangle_{C_1-C_2} \cdot \frac{d^2 \sigma_{\mbox{\scriptsize{\it{pp}}}}^{hard}}{dp_Tdy} 
\\
\frac{(d^2
\sigma_{\mbox{\tiny{\it{pA,AB}}}}^{\mbox{\scriptsize{\it{hard}}}})_{C_1-C_2}}{dp_Tdy} 
& = & \langle T_{A,AB}\rangle_{C_1-C_2} \cdot 
(\sigma_{\mbox{\tiny{\it{pA,AB}}}}^{geo})_{C_1-C_2} \cdot 
\frac{d^2\sigma_{\mbox{\scriptsize{\it{pp}}}}^{\mbox{\scriptsize{\it{hard}}}}}{dp_Tdy}
\\
\frac{(d^2 {\cal N}_{\mbox{\tiny{\it{pA,AB}}}}^{\mbox{\scriptsize{\it{hard}}}})_{C_1-C_2}}{dp_Tdy}  & = & 
{\cal L}_{int} \cdot \langle T_{A,AB}\rangle_{C_1-C_2} \cdot 
(\sigma_{\mbox{\tiny{\it{pA,AB}}}}^{geo})_{C_1-C_2} \cdot 
\frac{d^2 \sigma_{\mbox{\scriptsize{\it{pp}}}}^{\mbox{\scriptsize{\it{hard}}}}}{dp_Tdy}
\\
\nonumber\\
\nonumber\\
\frac{(d^2 N_{\mbox{\tiny{\it{pA,AB}}}}^{hard})_{C_1-C_2}}{dp_Tdy} & = &
\langle N_{coll}\rangle_{C_1-C_2} \cdot \frac{d^2 N_{\mbox{\scriptsize{\it{pp}}}}^{hard}}{dp_Tdy} 
\\
\frac{(d^2
\sigma_{\mbox{\tiny{\it{pA,AB}}}}^{\mbox{\scriptsize{\it{hard}}}})_{C_1-C_2}}{dp_Tdy} 
& = &  \langle N_{coll}\rangle_{C_1-C_2} \cdot
(\sigma_{\mbox{\tiny{\it{pA,AB}}}}^{geo})_{C_1-C_2}
\cdot \frac{d^2 N_{\mbox{\scriptsize{\it{pp}}}}^{\mbox{\scriptsize{\it{hard}}}}}{dp_Tdy}
\\
\frac{(d^2 {\cal N}_{\mbox{\tiny{\it{pA,AB}}}}^{\mbox{\scriptsize{\it{hard}}}})_{C_1-C_2}}{dp_Tdy}  & = & 
{\cal L}_{int}\cdot \langle N_{coll}\rangle_{C_1-C_2} \cdot 
(\sigma_{\mbox{\tiny{\it{pA,AB}}}}^{geo})_{C_1-C_2}
\cdot \frac{d^2 N_{\mbox{\scriptsize{\it{pp}}}}^{\mbox{\scriptsize{\it{hard}}}}}{dp_Tdy}
\end{eqnarray}

\vskip1cm
\noindent
\section*{Acknowledgments}
I would like to thank K. Reygers for the use of his Glauber Monte Carlo code and for useful comments,
H. Delagrange for valuable discussions, and A.~Morsch for detecting a few inconsistencies
in a previous version of the document. Support from the 5th European Union TMR Programme
(Marie-Curie Fellowship No. HPMF-CT-1999-00311) is also gratefully acknowledged.


\end{document}